\documentstyle[12pt,cite]{article}
\def\be{\begin{equation}}
\def\ee{\end{equation}}
\def\ba{\begin{eqnarray}}
\def\ea{\end{eqnarray}}
\def\ra{\rightarrow}
\def\bm#1{\mbox{$\boldmath{#1}$}}
\title{\bf Jarlskog-like invariants for theories with scalars
and fermions}
\author{F.\ J.\ Botella and Jo\~ao P.\ Silva \\
\small Departament de F\'{\i}sica Te\`{o}rica and IFIC\\
\small Universitat Val\`{e}ncia - CSIC\\
\small E-46100 Burjassot (Val\`{e}ncia), Spain}
\begin{document}
\maketitle
\begin{abstract}
Within the framework of theories where both scalars and fermions are
present, we develop a systematic prescription for the construction of
CP-violating quantities that are invariant under basis transformations
of those matter fields.
In theories with Spontaneous Symmetry Breaking,
the analysis involves the vevs' transformation properties under a scalar
basis change, with a considerable simplification of the study of
CP violation in the scalar sector.
These techniques are then applied in detail to the two Higgs-doublet
model with quarks.
It is shown that there are new invariants involving
scalar-fermion interactions, besides those already derived in previous
analyses for the fermion-gauge and scalar-gauge sectors.
\end{abstract}


\section{Introduction}

When a theory has several fields with the same quantum numbers,
one can rewrite the lagrangian in terms of new fields, gotten from
the original ones by a simple basis transformation. Obviously, the
result of a physical process does not depend on such redefinitions;
it can only depend on basis invariant quantities.
The construction of such basis invariant quantities is especially useful
in the context of CP violation, since the mere multiplication of a field
by a phase (rephasing) will originate spurious phases in the lagrangian.

This ``fuzziness'' of CP was already stressed in 1966 by Lee and Wick
\cite{leewick}, who pointed out that: {\it i)} CP is properly defined
for some portion of the total lagrangian ( ${\cal L}_G$ ) for which CP
is a good symmetry;
{\it ii)} any internal symmetry of ${\cal L}_G$ may be included in the
definition of CP. It has become customary to use for these the name of
Generalized CP (GCP) transformations. If the remaining portions of the
total lagrangian are manifestly invariant under one of these
GCP transformations, then the theory is CP conserving.

In the context of gauge theories, the standard procedure consists in
allowing for the inclusion of any weak basis transformation in the
definition of CP
\cite{jarlskogwu,gauge-fermion,menpom,lavsil,roldan}.
Here, the name of weak basis transformations
is abusively used to denote those transformations of the matter fields
that leave the gauge-matter interactions invariant, regardless of which
is the gauge group. One then uses the couplings of the remaining portions
of the lagrangian to build CP-violating, weak basis invariant quantities.
The simplest example arises in the SM with three families,
where there is only one independent CP-violating
basis invariant quantity \cite{jarlskogwu}, $J$,
arising in the complex Yukawa couplings
($M_u$ for the up-type quarks, and $M_d$ for the
down-type quarks),
\be
J \equiv \det [M_u M_u^{\dagger}, M_d M_d^{\dagger}]\ .
\label{eq:Jweak}
\ee
This quantity can be parametrized in terms of the Yukawa couplings
and charged gauge-fermion couplings written in the mass basis
-- $D_u = diag(m_u,m_c,m_t)$, $D_d = diag(m_d, m_s, m_b)$ and
$V$, respectively --
as
%
%
\footnote{Of course, one can break the lagrangian in a different way,
defining CP at the level of the kinetic and neutral gauge interactions.
These are invariant under general basis transformations, which, in
particular, may transform the left-handed fields $u_L$ and $d_L$
differently. One then uses the transformations of both the charged
gauge-fermion and the yukawa couplings to build the CP-violating
quantity
$\det [V^\dagger M_u M_u^{\dagger} V, M_d M_d^{\dagger}]$
which is explicitly invariant under {\it any} basis transformation.
In a weak basis ($V=1$) this expression reduces to Eq.~(\ref{eq:Jweak}),
while in the mass basis it reduces to Eq.~(\ref{eq:Jmass}).
The reason behind using transformations that leave the gauge-matter
interactions is apparent; if one does not, one must also consider
the transformation properties of the gauge-matter couplings,
significantly complicating the analysis.
}.
%
%
%
\ba
J & \equiv & \det [V^\dagger D_u D_u^{\dagger} V, D_d D_d^{\dagger}]
\nonumber\\
  & \propto &
(m_t^2 - m_c^2)
(m_t^2 - m_u^2)
(m_c^2 - m_u^2)
(m_b^2 - m_s^2)
(m_b^2 - m_d^2)
(m_s^2 - m_d^2)
\nonumber\\
  &  &
\times {\rm Im} (V_{ud} V_{cs} V_{us}^{\ast} V_{cd}^{\ast})\, ,
\label{eq:Jmass}
\ea
For historical reasons,
we shall use the name of Jarlskog-like invariants (or just $J$-invariants,
for simplicity) for all CP-violating basis invariant quantities.
The construction of $J$-invariants has been done in a variety of models.
The method of choice has been to cleverly look for quantities that
transform into themselves under a basis transformation but develop a
minus sign under a GCP transformation.
The fermion-gauge (-mass) sector was the first to be fully analyzed
in a series of models \cite{gauge-fermion}.
Similar $J$-invariants were later developed for the gauge-scalar sector
of multi Higgs-doublet models without fermions \cite{menpom,lavsil}.

In this article, we present the rules for the systematic construction
of $J$-invariants in any gauge theory with fermions and scalars including
the gauge-fermion, gauge-scalar but also the scalar-fermion sources of
CP-violation. In Section 2 we discuss an alternative method for the
construction of $J$-invariants inspired by perturbation theory.
This enables a much cleaner study of the CP-violation in the
gauge-scalar sector than that performed in Ref.~\cite{lavsil}.
In addition, it allows for the straightforward inclusion of fermions
in theories with many scalars.
The two Higgs-doublet model (2HDM) is worked out in detail
in Section 3 where new $J$-invariants signaling CP violation in
the scalar-gauge sector are identified.
In Section 4 we draw our conclusions.

\section{Systematic construction of $J$-invariants}

We will try to motivate our general prescription for the construction
of $J$-invariants with some examples. We start with a generic lagrangian of
the form,
\be
{\cal L}_I = g_{ij} \alpha_i \beta_j \Phi +
h_{kl} \alpha_k \gamma_l \Phi + h.c.\ ,
\ee
where $g$ and $h$ are coupling constants and $\alpha_i$, $\beta_i$ and
$\gamma_i$ are field operators defining the $i$th direction in the respective
U($\alpha$), U($\beta$) and U($\gamma$) flavour spaces, and
transforming like some multiplet of the gauge group, $G$.
As an example, for the three family SM we have, after SSB,
\be
{\cal L}_I
 =
( {\bar u}_L, {\bar d}_L)_i\  {M_u}_{ij}
\left( \begin{array}{c}
		1\\
		0
	\end{array}
\right){u_R}_j
+
( {\bar u}_L, {\bar d}_L)_i\  {M_d}_{ij}
\left( \begin{array}{c}
		0\\
		1
	\end{array}
\right){d_R}_j
+
h.c.\ ,
\ee
with the flavour spaces being U(3)$_L$, U(3)$_{uR}$ and U(3)$_{dR}$,
respectively. In perturbation theory, one can generate interactions
mediated by any power of ${\cal L}_I$. For example, to second order
in perturbation theory, we will find interactions mediated by
\be
(g_{ij} \alpha_i \beta_j \Phi)\
(h_{kl} \alpha_k \gamma_l \Phi)\ .
\ee
Hence, a given property of the theory (say CP violation) may show up at
some order of perturbation theory as a suitable product of couplings.

Under a basis transformation the couplings transform as,
\ba
g_{ij} & \rightarrow &
U(\alpha)_{ki}\  g_{kl}\  U(\beta)_{lj}\ ,
\nonumber\\
h_{ij} & \rightarrow &
U(\alpha)_{ki}\  h_{kl}\  U(\gamma)_{lj}\ .
\ea
The strategy in looking for basis invariant quantities consists in taking
products of couplings (as in the perturbative expansion),
contracting over the internal flavour spaces and taking a trace at the end.
For example, the quantities
\be
H_u = M_u M_u^\dagger\ \ ,\ \ H_d = M_d M_d^\dagger\ \ ,
\ \ H_u H_d
\ee
are tensors in the U(3)$_L$ space, whose traces are weak basis invariant.
The same is true for the trace of the U(3)$_{uR}$ tensor
$M_u^\dagger M_u$.

In so doing, we have already traced over the basis transformations
that could lead to the spurious phases that we alluded to in the
introduction.
Therefore, the imaginary parts of such traces are unequivocal
signs of CP violation \cite{roldan}.
For example, the three family $J$-invariant can be rederived as
\cite{roldan},
\be
J\ \propto\
{\rm Im}
\{
Tr
(H_u H_d H_u^2 H_d^2)
\}
\ .
\ee
This method has already been used to discover an $\epsilon$-type
contribution to baryogenesis in SU(5) with two Higgs fiveplets
\cite{botella}, in addition to
the $\epsilon^\prime$-type contribution found earlier \cite{nanopoulos}.

A final detail concerns spontaneous symmetry breaking (SSB).
After SSB, the physical degrees of freedom of the neutral scalars are
described by the shifted fields ($\eta_i$) related to the original ones
($\phi_i$) by the vevs ($v_i$) as,
\be
\phi_i  = v_i + \eta_i\ ,
\ee
which reparametrizes a lagrangian term such as
\be
\bm{ a}_{ij} \phi_i^\dagger \phi_j\ ,
\ee
into
\be
\bm{ a}_{ij} v_i^\ast v_j
+\bm{ a}_{ij} v_i^\ast \eta_j
+\bm{ a}_{ij} \eta_i^\dagger v_j
+\bm{ a}_{ij} \eta_i^\dagger \eta_j\ ,
\ee
where $v_i$ becomes an integral part of some new couplings.
Thus,
for the scalar sector the construction of the invariants must also include
the vacuum expectation values. As we will point out in the next section,
this greatly simplifies the study of the scalar sector over the previous
analysis of Ref.~\cite{lavsil}.
In addition, the minimization conditions provide relations between the
couplings in the scalar potential which must be used in identifying the
correct number of independent CP violating invariants.

This discussion motivates the following prescription for the construction
of $J$-invariants:
\begin{itemize}
\item identify all the scalar and fermion flavour spaces in the theory;
\item make a list of all the couplings according to their transformation
properties under weak basis transformations, including the vacuum
expectation values (which transform as vectors under the scalar basis
change), and make use of the stationarity conditions of the scalar
potential to reduce the number of parameters;
\item construct invariants by contracting over internal flavour spaces
in all possible ways, taking traces at the end (to be systematic
it is best to do this first in the fermion sector, say, and then use
this to define new scalar tensors, performing the scalar analysis as a
next step);
\item take the imaginary part to obtain a basis invariant signal of
CP violation.
\end{itemize}
Note that, in general, a minimal set of CP violating quantities is
not easy to find since one could in principle go to arbritrary order
in perturbation theory. That analysis is best done on a case by case
basis through a careful study of the CP violation sources in the model.
Moreover, different particular cases of a model may require different
choices for the fundamental $J$-invariants. This is the case even in
models as simple as the four family SM \cite{gronau}.

One should notice the generality of the proposed method.
In fact, this scheme applies to any gauge group, $G$.
In addition, there is no renormalizability requirement. The method
is applicable to any effective field theory with renormalizable,
as well as nonrenormalizable interactions.

\section{The two Higgs-doublet model with quarks}

We will now look at an SU(2)$\otimes$U(1) gauge theory with two
Higgs-doublets and with $n$ quark families. The particle content of
the theory consists of two scalar doublets
\be
\Phi_1\ \ \ ,\ \ \ \Phi_2\ ,
\ee
to which, without loss of generality, we can attribute the vevs
$v_1/\sqrt{2}$ and $v_2 e^{i \alpha}/\sqrt{2}$,
with $v_1$ and $v_2$ real, and the quark fields
\be
{\bar q}_L = ({\bar p}_L\ ,\ {\bar n}_L)\ \ ,\ \
p_R\ \ ,\ \ n_R\ ,
\ee
which are $n$-plets in the corresponding flavour spaces: respectively,
the spaces of SU(2) doublets, charged $2/3$ singlets and
charged $- 1/3$ singlets.
The Yukawa lagrangian is,
\be
- {\cal L}_Y =
{\bar q}_L \Gamma_i n_R \Phi_i +
{\bar q}_L \Delta_i^\ast p_R \tilde{\Phi}_i +
h.c.\ ,
\ee
where the $n \times n$ matrices $\Gamma_i$ and $\Delta_i^\ast$
%
%
\footnote{
The reason behind the noncanonical definition of
matrices $\Delta_i$ with the complex conjugation
will become apparent in
Eq.~(\ref{eq:transfT}).
}
%
%
contain the Yukawa couplings to the scalar $\Phi_i$,
and a sum over the scalar space ($i=1,2$) is implicit.
The scalar potential may be written as:
\be
V_H =
\bm{ a}_{ij} (\Phi_i^\dagger \Phi_j) +
\bm{ l}_{ij,kl} (\Phi_i^\dagger \Phi_j) (\Phi_k^\dagger \Phi_l)\ ,
\ee
where hermiticity implies
\ba
\bm{ a}_{ij}
& = &
\bm{ a}_{ji}^\ast\ ,
\nonumber\\
\bm{ l}_{ij,kl} \equiv \bm{ l}_{kl,ij}
& = &
\bm{ l}_{ji,lk}^\ast\ .
\ea
The stationarity conditions are
\be
v_i^\ast
\left[ \bm{ a}_{i \alpha}
+ 2 v_k^\ast\  \bm{ l}_{i \alpha,kl}\  v_l
\right] = 0
\hspace{10mm}
(\mbox{for}\  \alpha = 1, 2)\ .
\label{eq:statcond}
\ee
We have used boldfaced characters to remind us that these are tensors in
the scalar space.

Under weak basis redefinitions of scalars (through a unitary matrix $U$),
and of fermions (through matrices $X$), the Yukawa couplings get
transformed as \cite{vienna}
\ba
\Gamma_i
& \ra &
X_L^\dagger \Gamma_j U_{ji} X_{dR}\ ,
\nonumber\\
\Delta_i^\ast
& \ra &
X_L^\dagger \Delta_j^\ast U_{ji}^\ast X_{uR}\ ,
\label{eq:transfyuk}
\ea
with the scalar potential parameters and the vevs transforming as
\ba
\bm{ a}_{ij}   & \ra &
U_{ki}^\ast \bm{ a}_{kl} U_{lj}\ ,
\nonumber\\
\bm{ l}_{ij,kl}   & \ra &
U_{mi}^\ast U_{ok}^\ast \bm{ l}_{mn,op} U_{nj} U_{pl}\ ,
\label{eq:transfsca}\\
v_i & \ra &
U_{ji}^\ast v_j
\label{eq:transfvev}\ .
\ea
Following our general scheme, we first build fermion basis invariants
such as the traces
\ba
\bm{ T}_{ij}^\Gamma   & = &
Tr^f (\Gamma_i \Gamma_j^\dagger)
\nonumber\\
\bm{ T}_{ij}^\Delta   & = &
Tr^f (\Delta_i \Delta_j^\dagger)
\label{eq:Ttensors}
\ea
where $Tr^f$ indicates that a trace has been taken over the relevant
fermion flavour space.
These tensors, which are second order in the Yukawa couplings,
transform under a scalar basis change as
\be
\bm{ T}_{ij} \ra U_{ki}^\ast \bm{ T}_{kl} U_{lj}\ .
\label{eq:transfT}
\ee
One may now combine the scalar basis tensors in
Eqs.~(\ref{eq:transfsca}), (\ref{eq:transfvev}) and
(\ref{eq:Ttensors}), taking a trace at the end and thus obtaining
weak basis independent quantities.
If these have a nonzero imaginary part, we will have a sign of
CP violation.

Being the only vector, $v_i$ must always appear in the scalar traces
in the combination
\be
\bm{ V}_{ij} = v_i v_j^\ast\ .
\ee
Further, all these second rank tensors are hermitian and hence one needs
either three different ones or the repetition of two of them in
order to get a complex trace. For example
\be
Tr( \bm{ V} \bm{ a})\ , \hspace{10mm}
Tr( \bm{ V}^2 \bm{ a})\ ,
\ee
are clearly real, while
\be
J_a = {\rm Im}\ Tr( \bm{ V} \bm{ a} \bm{ T}^\Gamma)\ , \hspace{10mm}
J_b = {\rm Im}\ Tr( \bm{ V} \bm{ a} \bm{ T}^\Delta)\ ,
\label{eq:Jalpha}
\ee
which mix the scalar and fermion sectors, or
\be
J_1 = {\rm Im}\ (v_i^\ast v_j^\ast \bm{ a}_{i \alpha} \bm{ a}_{j \beta}
\bm{ l}_{\alpha k, \beta l} v_k v_l)\ ,
\hspace{10mm}
J_3 = {\rm Im}\ (v_i^\ast \bm{ a}_{ij}
\bm{ l}_{jk,kl} v_l)
\label{eq:Ji}\ ,
\ee
which depend exclusively on the scalar sector, may be nonzero.

In this form the invariants are difficult to interpret and,
moreover, it is not always clear how to include in them the
stationarity conditions (\ref{eq:statcond}).
A much clearer picture arises if one transforms the scalars into a
basis in which only the first scalar has a vacuum expectation value.

\subsection{The Higgs basis}

Indeed, since in this model all the scalars are in the same representation
of SU(2), the vev may be rotated all into the first scalar through the
transformation
\be
\Phi_i = U_{ij} H_j
\ee
with
\be
U^\dagger = \frac{1}{v}
\left( \begin{array}{cc}
	v_1 & v_2 e^{- i \alpha}\\
	v_2 & - v_1 e^{- i \alpha}
	\end{array}
\right)
\label{eq:U}\ .
\ee
Therefore, the new scalars may be parametrized as
\be
H_1  =
\left( \begin{array}{c}
	G^+ \\
	(v + H^0 + i G^0)/\sqrt{2}
	\end{array}
\right)\,
\hspace{10mm}
H_2  =
\left( \begin{array}{c}
	H^+ \\
	(R + i I)/\sqrt{2}
	\end{array}
\right)\, ,
\label{eq:Higgs}
\ee
where $ G^+ $ and $ G^0 $ are the Goldstone bosons,
which,
in the unitary gauge,
become the longitudinal components of the $ W^+ $ and of the $ Z^0 $,
and $ H^0 $, $ R $ and $ I $ are real neutral fields, with $H^0$
coupling to fermions proportionally to their masses (in the fermion
mass basis). Note that these features remain the same if one multiplies
$H_2$ by a phase. All that does is to rotate $R$ and $I$ through an
orthogonal transformation.

In this basis, the Yukawa coupling matrices become
\be
\Gamma_i^{H} = \Gamma_j U_{ji}\ ,
\hspace{10mm}
\Delta_i^{H \ast} = \Delta_j^\ast U_{ji}^\ast\ ,
\ee
with the scalar couplings transformed into $\bm{ \mu}_{ij}$
and $\bm{ \lambda}_{ij,kl}$,  given in terms of the original ones
by the right hand side of Eq.~(\ref{eq:transfsca}), with the specific
form of $U$ written in Eq.~(\ref{eq:U}).
With an obvious change of notation ($\bm{ \mu}_{11} = \mu_1$,
$\bm{ \mu}_{12} = \mu_3$, etc.) we can write the scalar potential
in the familiar form
\ba
V_H & = &
\mu_1 H_1^{\dagger} H_1 + \mu_2 H_2^{\dagger} H_2
+ (\mu_3 H_1^{\dagger} H_2 + h.c.)
\nonumber\\
  &   &
+ \lambda_1 (H_1^{\dagger} H_1)^2
+ \lambda_2 (H_2^{\dagger} H_2)^2
+ \lambda_3 (H_1^{\dagger} H_1) (H_2^{\dagger} H_2)
+ \lambda_4 (H_1^{\dagger} H_2) (H_2^{\dagger} H_1)
\nonumber\\
  &   &
+ \left[ \lambda_5 (H_1^{\dagger} H_2)^2
+ \lambda_6 (H_1^{\dagger} H_1) (H_1^{\dagger} H_2)
+ \lambda_7 (H_2^{\dagger} H_2) (H_1^{\dagger} H_2)
+ h.c.\right]\, ,
\label{eq:potential}
\ea
in which all the coupling constants,
except $ \mu_3 $,
$ \lambda_5 $,
$ \lambda_6 $,
and $ \lambda_7 $,
are real by hermiticity.

In this basis, the stability conditions of the vacuum
in Eq.~(\ref{eq:statcond}) take the very simple form
\be
\mu_1 = - \lambda_1 v^2\ ,
\hspace{10mm}
\mu_3 = - \lambda_6 v^2/2\ .
\ee
Using this last equation one can eliminate $\mu_3$ so that only
the phases of $\lambda_5$, $\lambda_6$ and $\lambda_7$ remain.
The simplicity of these relations, and the fact that only $H_1$ has
a vev, greatly simplify the form of the invariants.
For example, in this basis the invariants of Eq.~(\ref{eq:Ji})
take the very simple form \cite{lavsil}
\be
J_1 \propto {\rm Im} (\lambda_6^2 \lambda_5^\ast)\ ,
\hspace{10mm}
J_3 \propto {\rm Im} (\lambda_6 \lambda_7^\ast)\ .
\ee
These are the only two independent CP-violating basis invariant quantities
of the most general 2HDM without fermions.
The other possible invariant,
\be
J_2 \propto {\rm Im} (\lambda_7^2 \lambda_5^\ast)\ ,
\ee
is just a combination of the previous two.

These quantities were derived in Ref.~\cite{lavsil}
by looking for CP-violating quantities, invariant under a phase
transformation of $H_2$.
This required the expression of $V_H$ in terms of the
component field of Eq.~(\ref{eq:Higgs}), and the subsequent analysis
of the transformation properties of each term in
$V_H$ under that rephasing, a procedure that is rather tedious.
With the methods we have developed above, this analysis becomes
straightforward, mainly due to the inclusion of the vevs' transformation
properties. Further, it becomes clear that these are invariants under
any weak basis change, and not just under a rephasing of $H_2$.
They just happen to be written in the Higgs basis for simplicity.

\subsection{The quark mass basis}

As happens in the SM for the Jarlskog invariant, the physical
interpretation of invariants involving fermions is clearer when
the Yukawa matrices are parametrized in terms of their values in
the quark mass basis.
This is done through transformations that diagonalize the couplings
of the quarks to the Higgs ($H_1$) with the vev $v/\sqrt{2}$:
\ba
v/\sqrt{2}\ {X_d^\dagger}_L \Gamma_1^{H} {X_d}_R
& = &
D_d = diag (m_d, m_s, \ldots )\ ,
\nonumber\\
v/\sqrt{2}\ {X_u^\dagger}_L \Delta_1^{H \ast} {X_u}_R
& = &
D_u = diag (m_u, m_c, \ldots )\ ,
\ea
while the couplings to the vevless Higgs ($H_2$) become
\ba
v/\sqrt{2}\ {X_d^\dagger}_L \Gamma_2^{H} {X_d}_R
& = &
N_d\ ,
\nonumber\\
v/\sqrt{2}\ {X_u^\dagger}_L \Delta_2^{H \ast} {X_u}_R
& = &
N_u\ ,
\ea
where $V = {X_u^\dagger}_L {X_d}_L$ is the CKM matrix.
The Yukawa lagrangian may then be written as
\ba
- {\cal L}_Y & = &
\left({ {\bar u}_L V\ , {\bar d}_L }\right)
\left[{ D_d \left({ \frac{\sqrt{2}}{v} H_1 }\right)
+N_d \left({ \frac{\sqrt{2}}{v} H_2 }\right) }\right]\
d_R
\nonumber\\
    & + &
\left({ {\bar u}_L\ , {\bar d}_L V^\dagger }\right)
\left[{ D_u \left({ \frac{\sqrt{2}}{v} \tilde{H_1} }\right)
+N_u \left({ \frac{\sqrt{2}}{v} \tilde{H_2} }\right) }\right]\
u_R +h.c.\ .
\label{eq:yukhiggs}
\ea
One can still perform equal rephasings on the left- and right-handed
components of each quark without affecting these properties.
Usually this is used to remove $2n-1$ unphysical phases from the CKM
matrix. The matrices $N_d$ and $N_u$ are, however, perfectly
arbritrary complex $n \times n$ matrices.

In this basis the fermion traces of Eq.~(\ref{eq:Ttensors}) are
\ba
\frac{v^2}{2} \bm{T}_{11}^\Gamma & = &
\sum_{i=1}^{n} {m_d^2}_i\ ,
\nonumber\\
\frac{v^2}{2} \bm{T}_{22}^\Gamma & = &
\sum_{i,k=1}^{n} |{N_d}_{ik}|^2\ ,
\nonumber\\
\frac{v^2}{2} \bm{T}_{12}^\Gamma  \equiv
\frac{v^2}{2} \bm{T}_{21}^{\Gamma \ast} & = &
\sum_{i=1}^{n} {m_d}_i {N_d}_{ii}\ ,
\ea
with similar expression for the up sector.
The invariants of Eq.~(\ref{eq:Jalpha}) thus become
\ba
J_a & = &
{\rm Im} (\bm{\mu}_{12} \bm{T}_{21}^\Gamma)
\nonumber\\
   & = &
\sum_{i=1}^{n} {\rm Im} ({m_d}_i\  \mu_3\  {N_d}^\ast_{ii})\ ,
\nonumber\\
J_b & = &
\sum_{i=1}^{n} {\rm Im} ({m_u}_i^\ast\  \mu_3\  {N_u}_{ii})\ .
\label{eq:Jalphamass}
\ea
At first sight this is a surprising result since only one power
of the mass is involved, contrary to our SM acquired intuition.
The point is that, in the SM there is only one Yukawa matrix for
the down-type quarks, say, and therefore it needs to appear twice
to preclude changes in the invariants arising from different
rephasings of the right and left-handed components of some quark.
Here, as is seen explicitly in Eq.~(\ref{eq:Jalphamass}),
the existence of two matrices allows for the construction
of invariants in which the phase change in one of those matrices
is compensated by the same phase change in the other.
To be more specific, imagine that we perform
a phase change on the $d_R$ quark,
\be
d_R \ra e^{i \delta} d_R.
\ee
Then,
\ba
{D_d}_{11} & \ra & {D_d}_{11} e^{i \delta}\ ,
\nonumber\\
{N_d}_{11} & \ra & {N_d}_{11} e^{i \delta}\ ,
\ea
and their rephasings get cancelled in the Eq.~(\ref{eq:Jalphamass}).
It was to emphasize this point that the masses (which are real positive
numbers in the mass basis) were kept inside the imaginary part in these
equations.

The appearance of $\mu_3$ (or better, $\lambda_6$, once the stationarity
conditions are used) is also easy to understand.
If, for example, one rephases $H_2$ by
\be
H_2 \ra e^{i \xi} H_2\ ,
\ee
one gets
\ba
\mu_3 & \ra & \mu_3 e^{i \xi}\ ,
\nonumber\\
N_d & \ra & N_d e^{i \xi}\ ,
\nonumber\\
N_u & \ra & N_u e^{-i \xi}\ ,
\ea
and hence $\mu_3$ must appear in combination with $N_d^\dagger$
(or $N_u$).
A similar rephasing of $H_1$ leads to the conclusion that
$D_d$ (or $D_u^\dagger$) must also be involved.
We have thus succeeded in constructing weak basis invariant
quantities that control the feeding of phases between the fermion
and the scalar sectors, through the rephasings of either quarks or scalars.

For the simplest case of just one quark family, it is easy to see that
we have all the invariants we need. In fact, in the mass basis, there
are five complex quantities:
$\lambda_5$, $\lambda_6$ and $\lambda_7$ in the scalar sector
($\mu_3$ is related to $\lambda_6$ through the stationarity
conditions);
and $N_d$ and $N_u$ in the Yukawa couplings, which are now just
complex numbers. The freedom to rephase $H_2$ allows us to set
one of these quantities real and only four $J$-invariants remain.
We can take these to be $J_1$, $J_3$, $J_a$ and $J_b$.
However, depending on the particular case in question, other
combinations might be more useful. For instance,
in a model with $\mu_3 = 0 = \lambda_6$, all of the above are zero
but there are still three CP-violating phases.
One is $J_2$ and the others may be chosen as
\begin{eqnarray}
J_x & = &
{\rm Im} (m_u m_d N_d^\ast N_u^\ast)\ ,
\label{Jx}\\
J_y & = &
{\rm Im} (m_d \lambda_7 N_d^\ast)\ .
\label{Jy}
\end{eqnarray}
The invariant $J_x$ appears for example in the charged Higgs boson
contribution to the decay rate asymmetry
$\Gamma[{\bar b} \ra {\bar s} \gamma]
-\Gamma[b \ra s \gamma]$
and to the dipole moment of the neutron,
in the approximation in which the third quark family decouples from
the first two \cite{ylwu}.

The situation is similar for more quark families.
In general one has $J_1$ and $J_3$ from the scalar sector,
and one needs to find an extra $2 n^2$
(from the phases in the $N$ matrices) plus $(n-1)(n-2)/2$
(from the irremovable phases in the CKM matrix) invariants,
using the methods described above.
As a simple illustration we can look at the 2HDM with
two families. There are now 8 phases in the couplings to $H_2$ and
yet no phase in the CKM matrix (in the parametrization we described above).
Besides the fermion traces of Eq.~(\ref{eq:Ttensors}), we now need
also the tensors
\ba
\bm{T}^{\Gamma \Gamma}_{ijkl} & = &
Tr (\Gamma_i \Gamma^\dagger_j \Gamma_k \Gamma^\dagger_l)\ ,
\nonumber\\
\bm{T}^{\Delta \Delta}_{ijkl} & = &
Tr (\Delta_i \Delta^\dagger_j \Delta_k \Delta^\dagger_l)\ ,
\label{eq:TTtensors}
\ea
and the quantities
\ba
\bm{T}^{\Gamma \Delta}_{ijkl} & = &
Tr (\Gamma_i \Gamma^\dagger_j \Delta_k^{\ast} \Delta^{\ast \dagger}_l)\ ,
\label{eq:TTmixedtensor}
\ea
which transforms in the last two indices as $U^\ast$ and
$U^\top$, because of the required inclusion of $\Delta^\ast$
to compensate for the transformation properties of $\Gamma$ under the
quark left-handed flavour transformations. Of course, one must keep this
in mind when performing the scalar traces. It is easy to check that
the phases ${N_d}_{11}$ and ${N_d}_{22}$ are directly related to
$\bm{T}^{\Gamma}_{12}$ and $\bm{T}^{\Gamma \Gamma}_{1112}$.
The phase of ${N_d}_{12} {N_d}_{21}$ can then be easily related to
$\bm{T}^{\Gamma \Gamma}_{1212}$ or $\bm{T}^{\Gamma \Gamma}_{2212}$.
Similarly for the up quark sector. The final two independent
combinations of phases can be related, for example, to
\ba
\bm{T}^{\Gamma \Delta}_{1112} & = &
Tr (D_d D_d^\dagger V^\dagger D_u N_u^\dagger V)
\nonumber\\
  & = & m_d^2 m_c {N_u}_{22}^\ast + m_s^2 m_u {N_u}_{11}^\ast
\nonumber\\
   & + &
(m_s^2 - m_d^2) c_\theta
\left[ m_c ({N_u}_{22}^\ast + s_\theta {N_u}_{12}^\ast)
- m_u ({N_u}_{11}^\ast - s_\theta {N_u}_{21}^\ast) \right]\ ,
\nonumber\\
\bm{T}^{\Gamma \Delta}_{1211} & = &
Tr (D_d N_d^\dagger V^\dagger D_u D_u^\dagger V)
\nonumber\\
  & = &
m_u^2 m_s {N_d}_{22}^\ast + m_c^2 m_d {N_d}_{11}^\ast
\nonumber\\
   & + &
(m_c^2 - m_u^2) c_\theta
\left[ m_s ({N_d}_{22}^\ast - s_\theta {N_d}_{12}^\ast)
- m_d ({N_d}_{11}^\ast + s_\theta {N_d}_{21}^\ast) \right]\ ,
\label{eq:last2phases}
\ea
where the mass basis parametrization was used, and $s_\theta$
and $c_\theta$ are the sine and cosine of the Cabbibo angle,
respectively.
Clearly the $J_a$ and $J_b$ of Eq.~(\ref{eq:Jalphamass})
are still invariants. In addition, 6 new invariants can be made out
in the same way as in
Eq.~(\ref{eq:Ji}), substituting $\bm{l}_{ijkl}$ for
the tensors of Eqs.~(\ref{eq:TTtensors}) and
(\ref{eq:TTmixedtensor}). These have interpretations similar
to the simpler ones of the one family, 2HDM above.
The game to be played for three families is now obvious, with the
added feature that in that case the CKM matrix has an irremovable phase
leading to the original invariant: the gauge-fermion (-mass)
Jarlskog invariant \cite{jarlskogwu}.

Whether these invariants are equal to zero or proportional to each other
depends on the model one considers. In the very popular 2HDM with
soft breaking of the $Z_2$ symmetry and spontaneous CP breaking
\cite{soft},
there is only one CP violating phase: the relative phase between
the two vacuua. Then, the Jarlskog invariant is zero,
and all other invariants are, of course, either zero or
proportional to $\sin{\alpha}$.

\section{Conclusion}

We have presented a systematic method for the construction of
basis invariant CP-violating quantities for theories with both
scalars and fermions.
This procedure was inspired by perturbation
theory and will reproduce the CP violations occurring perturbatively.
We also point out that, due to SSB, one must take the into account the
transformation properties of the vevs,
and have shown how that leads to a very simple analysis of the
CP violation in the scalar sector.

A simple application of this scheme was worked out in detail
for the 2HDM with $n$ quark families, for which we
reproduced the earlier results for the CP-violating
invariants in the gauge-fermion and gauge-scalar.
In addition, we have constructed new invariants that express the
CP violation in the scalar-fermion sector.

\vspace{5mm}

We thank J.\ Bernabeu, A.\ S.\ Joshipura and L.\ Wolfenstein for many
useful discussions and for critical readings of this manuscript.
We also thank A.\ Pich for various discussions.
This work has been supported in part by CICYT under grant AEN 93-0234.
The work of J.\ P.\ S.\ was funded by the E.\ U.\ under the program
of Human Capital and Mobility.
J.\ P.\ S.\ is indebted to the Centro de F\'{\i}sica Nuclear of the
University of Lisbon and to Prof. A.\ M.\ Eir\'{o}
for their hospitality during the initial stages of this work.


\vspace{5mm}

\end{document}